\documentclass[aps,prb,preprint,superscriptaddress]{revtex4}
\usepackage{graphicx}
\usepackage{amsmath}

\begin{document}
\title{Charge mobility determination by current extraction under linear
  increasing voltages: the case of non-equilibrium charges and field-dependent mobilities}
\author{Sebastian Bange}
\email{sbange@uni-potsdam.de}
\author{Marcel Schubert}
\author{Dieter Neher}
\affiliation{Universit\"{a}t Potsdam, Institut f\"{u}r Physik und Astronomie, Karl-Liebknecht-Str.~24--25, 14476 Potsdam-Golm, Germany}
\date{\today}

\begin{abstract}
  The method of current extraction under linear increasing voltages (CELIV)
  allows for the simultaneous determination of charge mobilities and charge
  densities directly in thin films as used in organic photovoltaic cells
  (OPV). In the past, it has been specifically applied to investigate the
  interrelation of microstructure and charge transport properties in such
  systems. Numerical and analytical calculations presented in this work show
  that the evaluation of CELIV transients with the commonly used analysis
  scheme is error prone once charge recombination and, possibly, field
  dependent charge mobilities are taken into account. The most important
  effects are an apparent time-dependence of charge mobilities and errors in
  the determined field dependencies. Our results implicate that reports on
  time-dependent mobility relaxation in OPV materials obtained by the CELIV
  technique should be carefully revisited and confirmed by other measurement
  methods.
\end{abstract}
\maketitle

\section{Introduction}
\label{sec:introduction}

The understandig of charge transport properties of organic semiconducting
materials is one of the keys for further improvement of devices such as
organic light-emitting diodes and organic photovoltaic cells. Owned to it's
simplicity, straightforward data analysis and applicability to measurements on
films of well below 100~nm thickness, the technique of current extraction by
linear increasing voltages (CELIV, see Ref.~\onlinecite{Juska2000}) has attracted
considerable interest over the past few years. It's unique attractiveness
stems from the opportunity to study charge transport directly in thin film
systems as used in the actual devices. As such, it is the ideal technique to
investigate structure-property relationships of charge transport and
recombination e.g. in state-of-the-art donor/acceptor photovoltaic systems
such as polymer/polymer and polymer/small molecule material blends, the
morphology of which cannot be well reproduced in modified sample
geometries. CELIV has been originally introduced to determine charge mobility
and concentration in microcrystalline Si:H semiconductors and doped conjugated
polymers under thermal equilibrium conditions,\cite{Juska2000,Juska2000b} but
was later applied to study the field-dependent mobility of non-equilibrium
charge carriers created by the absorption of a light
pulse.\cite{Juska2003,Genevicius2003,Mozer2005} Hereby photo-CELIV experiments
have been used to quantify the decay of carrier density with time and to
measure the time dependence of mobility. Time-dependent relaxation of charge
mobilities as determined by the photo-CELIV technique has been reported for
various state-of-the-art photovoltaic bulk heterojunction
blends.\cite{Osterbacka2004,Mozer2005c,Yin2008,Schubert2009,Homa2009} It has
usually been attributed to the relaxation of carrier energies in an extended
density of transport states present in these disordered solids.

This paper treats the analytic and numerical analysis of CELIV experiments
under conditions diverging from the equilibrium assumptions of its original
derivation. We first review the basic theory of equilibrium CELIV experiments
and point out inaccuracies in the original derivations. Based on this, we
treat the case of non-equilibrium photogenerated charge carriers with constant
mobility by both numerical as well as analytical methods. We show that with
the standard evaluation scheme applied under this condition, an apparently
time-dependent charge mobility is deduced that is solely the effect of the
used analysis. We provide an analytic analyis method to determine charge
mobilities in a reliable way even under conditions of non-equilibrium charge
carrier recombination. Finally, we treat the application of CELIV measurements
to the situation of field-dependent carrier mobilities. A modified analysis
method is introduced and shown to result in significantly reduced errors for the
derived field dependence.

\section{Charge mobility determination by the CELIV technique}
\label{sec:results-discussion}
\begin{figure}
  \centering
  \includegraphics{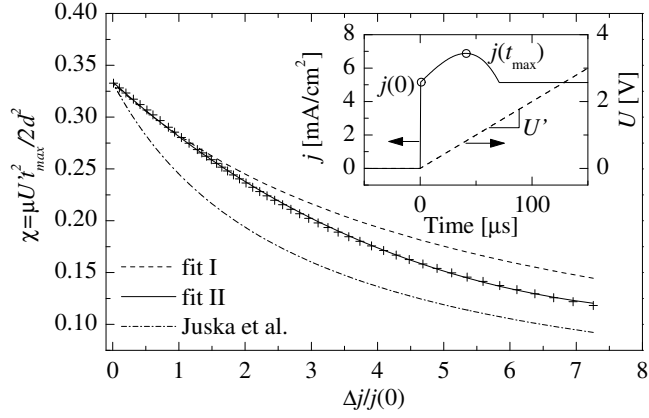}
  \caption{Results of the calculation of $\chi$ as function of $\Delta j/j(0)$
  from the numerical solution of equations~(\ref{eq:1}) and (\ref{eq:2})
  (symbols), compared to fit I and II
  as discussed in the text as well as the parametrisation used by Juska et
  al.\ in Ref.~\onlinecite{Juska2000b}. The inset shows a typical CELIV current calculated
for $d=100$~nm, $U'=2\times10^4$~V/s, $\mu=2\times10^6$~cm$^2$/Vs,
$n=10^{22}$~m$^{-3}$, $A=1$~mm$^2$ and $\epsilon_\text{r}=2.9$, indicating the
capacitive charging current $j(0)$ and the maximum current $j(t_\text{max})$.}
  \label{fig:Uptmaxfit}
\end{figure}

\begin{figure}
  \centering
  \includegraphics{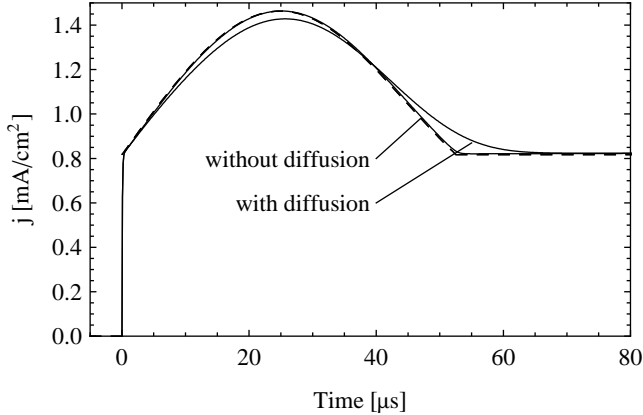}
  \caption{Numerically simulated CELIV transients (solid lines), for
    simulation parameters see text. The results are shown both with and
    without charge diffusion according to the Einstein relation. The dashed
    line corresonds to the numerical solution of equations (\ref{eq:1}) and
    (\ref{eq:2}) and closely follows the simulation results excluding
    charge diffusion.}
  \label{fig:Cmp-Numerical}
\end{figure}

The experimental procedure and analytic evaluation for CELIV experiments has
been introduced by Juska et al.\ in Ref.~\onlinecite{Juska2000}, to
which the reader is referred for further details. Consider the extraction of
equilibrium charge carriers of density $n$ and mobility $\mu$ in the electric
field $U/d$, where $d$ is the layer thickness and $U(t)=U't$ is the applied
voltage that rises linearly in time. Without restriction of the generality of
our analysis, we assume that the internal electric field is zero at
$U=0$. This differs from the more general situation where electrodes of
different work functions are used.  Note that experimentally one should
account for the built-in potential due to differences in the electrode work
functions to ensure that the electric field within the organic layer is zero
at $U=0$. Assuming that the electrode area $A$ is much larger than $d$, that
one carrier type is much more mobile than the other (here: holes) and that the
electrodes are non-injecting, the time dependent charge density is
$\rho(z,t)=-en$ for $0\leq z\leq l(t)$ and $\rho(z,t)=0$ for
$l(t)<z<d$. Hereby, all holes have been depleted from the layer up to the
time-dependent extraction depth $l(t)$. The current density measured in the
external circuit due to the extraction of charges at $z=d$ is
\begin{equation}
  \label{eq:1}
  j=\frac{\epsilon}{d}U'+\begin{cases}
    \frac{en}{d}\left(1-\frac{l}{d}\right)\left(\mu U't-\frac{en\mu
        l^2}{2\epsilon}\right)& (l(t)\leq d)\\ 0 & \text{else},\end{cases}
\end{equation}
assuming $t\gg RC$ where $R$ is the external circuit resistance and
$C=\epsilon A/d$ is the geometrical sample capacitance assuming a permittivity
of $\epsilon=\epsilon_\text{r}\epsilon_0$. The extraction depth is the solution of
\begin{equation}
  \label{eq:2}
  \frac{dl(t)}{dt}+\frac{en\mu l^2(t)}{2d\epsilon}=\frac{\mu U't}{d}
\end{equation}
under the initial conditions $l(0)=0$ and $dl(0)/dt=0$. This is a first order
nonlinear differential equation of Ricatti type that can be solved numerically
parametric in the dimensionsless voltage slope $\epsilon^2U'/2e^2n^2\mu
d^2$. At some time $t_\text{max}$, the current density~(\ref{eq:1}) will peak
at $j(t_\text{max})$, where the relative height $\Delta
j/j=(j(t_\text{max})-j(0))/j(0)$ can be expressed as a bijective function of
the dimensionless parameter $\chi=\mu
U't_\text{max}^2/2d^2$. Figure~\ref{fig:Uptmaxfit} shows the result of our
numerical calculation of $\chi$ as a function of $\Delta j/j(0)$. $\chi$ is
equal to 1/3 at $\Delta j/j(0)=0$ and decreases in a nonlinear fashion for
$\Delta j/j(0)>0$. Also shown in figure~\ref{fig:Uptmaxfit} are two different
parametrizations of this curve in terms of $\chi=[3(1+0.18\Delta
j/j(0))]^{-1}$ (fit I) and $\chi=0.329\exp[-0.180\Delta
j/j(0)]+0.005\exp[0.253\Delta j/j(0)]$ (fit II) which are good approximations
for $\Delta j/j(0)\leq1$ or $\Delta j/j(0)\leq7$ for fit I and fit II,
respectively.
Using these, the charge mobility can be
calculated as $\mu=2d^2\chi/U't_\text{max}^2$, resulting in
\begin{equation}
  \label{eq:6}
  \mu=\frac{2d^2}{3U't_\text{max}^2(1+0.18\frac{\Delta
      j}{j(0)})}  
\end{equation}
when using fit I valid for $\Delta j/j(0)\leq1$ or
\begin{equation}
  \label{eq:3}
    \mu=\frac{2d^2}{U't_\text{max}^2}\left(0.329e^{-0.180\frac{\Delta j}{j(0)}}+0.005e^{0.253\frac{\Delta j}{j(0)}}\right)
\end{equation}
when using fit II valid for $\Delta j/j(0)\leq7$. This is in variance with the
result published by Juska et al.\cite{Juska2000b,Juska2001} and used in several
articles,\cite{Juska2004,Osterbacka2004,Mozer2005b} which corresponds to
choosing $\chi=[3(1+0.36\Delta j/j(0))]^{-1}$. In order to provide an
independent test for the consistency of our analysis, we simulated the CELIV
experiment with a numerical drift-diffusion solver
program. Figure~\ref{fig:Cmp-Numerical} shows the resulting current transients
assuming a charge mobility of $\mu=2\times10^{-6}$~cm$^2$/Vs, a layer
thickness of $d=65$~nm with $\epsilon_\text{r}=3$, a voltage slope of
$U'=2\times10^4$~V/s and an initial charge carrier density of
$n=4\times10^{22}$~m$^{-3}$.
As expected, the numerically evaluated solution to equations~(\ref{eq:1}) and
(\ref{eq:2}) closely follows the simulation results when charge diffusion is
suppressed. We determined $\Delta j/j(0)=0.781$ and
$t_\text{max}=25.125$~$\mu$s from the simulation data, from which the apparent
mobilities $\mu=1.96\times10^6$~cm$^2$/Vs and $\mu=1.95\times10^6$~cm$^2$/Vs
are calculated using equation~(\ref{eq:6}) or equation~(\ref{eq:3}), respectively. Using the Juska et al.\
result, we instead obtaine $\mu=1.74\times10^{-6}$~cm$^2$/Vs, which proves
that this approximation underestimates the mobility. We therefore suggest to calculate the mobility
using equation~(\ref{eq:6}) or (\ref{eq:3}), depending on the magnitude of $\Delta j/j(0)$.

\section{The role of bimolecular charge recombination}
\begin{figure}
  \centering
  \includegraphics{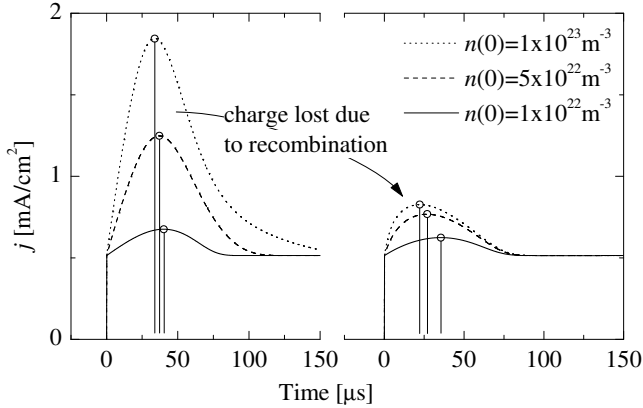}
  \caption{Comparison of numerically simulated CELIV current transients
    without (left) and with (right) bimolecular charge recombination according
    to the Langevin mechanism, for simulation parameters see text. The results
    are shown parametric in the initial charge density, the position of
    $j(t_\text{max})$ is marked by a circle and vertical line.}
  \label{fig:RecombinationSimulation}
\end{figure}

\begin{figure}
  \centering
  \includegraphics{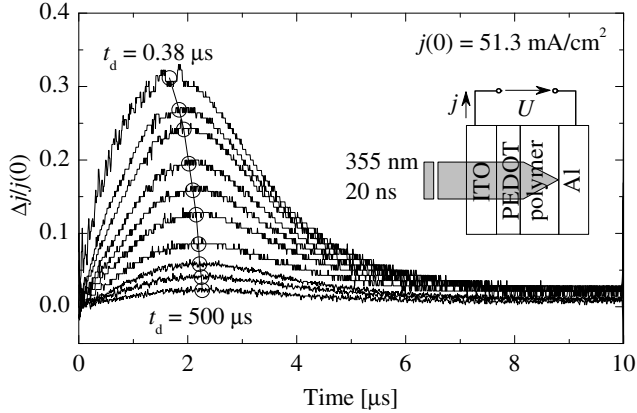}
  \caption{CELIV current transients measured for a 55~nm thick M3EH-PPV:CN-ether-PPV blend
  layer at $U'=1.06$~V/$\mu$s for different delay times after photoexcitation
  ranging from near zero up to 500~$\mu$s. Connected circles indicate the determined
  $j(t_\text{max})$ points. The inset schematically shows the layer structure
  and illumination direction used for the experiment.}
  \label{fig:CELIVTransientsExperiment}
\end{figure}

\begin{figure}
  \centering
  \includegraphics{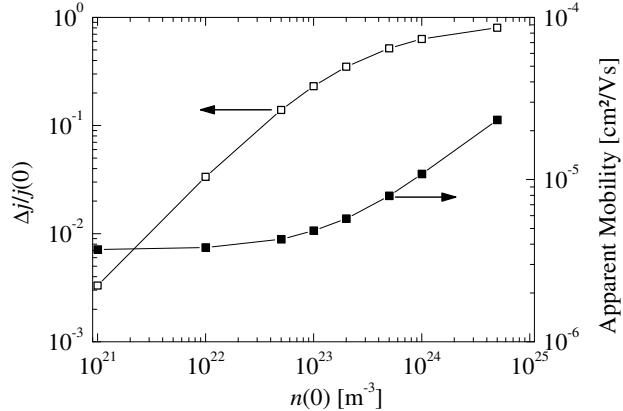}
  \caption{Analysis of numerically simulated CELIV transients in terms of
    current maximum $\Delta j/j(0)$ (open symbols) and apparent mobility
    (solid symbols) calculated from equation~(\ref{eq:3}) for a range of
    charge densities $n(0)$ present at the beginning of charge extraction
    ($t_\text{d}=0$). The simulation model parameters correspond to the
    experimental situation of figure~\ref{fig:CELIVTransientsExperiment}.}
  \label{fig:RecombinationSimulationII}
\end{figure}

\begin{figure}
  \centering
  \includegraphics{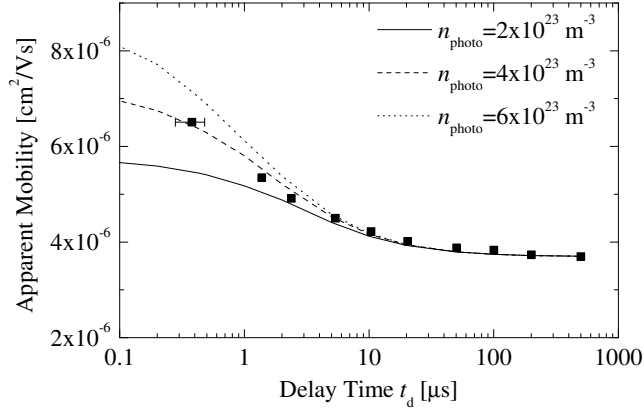}
  \caption{Apparent mobility calculated from the experimental CELIV transients
    shown in figure~\ref{fig:CELIVTransientsExperiment} using
    equation~(\ref{eq:3}) (symbols), compared with the results of numerical
    simulations assuming various photogenerated charge carrier densities
    $n_\text{photo}$, a time-\emph{independent} charge mobility
    $\mu=3.8\times10^{-6}$~cm$^2$/Vs and bimolecular recombination according
    to the Langevin mechanism. Data is shown as a function of the delay time
    $t_\text{d}$ between photoexcitation and the beginning of the extraction
    voltage pulse, the approximate uncertainty of delay time for first data
    point is indicated by error bars.}
  \label{fig:CELIVExperimentEvaluation}
\end{figure}

The gaussian disorder model of charge transport in organic semiconductors
predicts that after photoexcitation, charge carriers will relax energetically
towards their equilibrium energy, with a concomitant mobility decrease. The
understanding of this process is of considerable importance for organic
photovoltaic devices in order to further improve their efficiency. Recently,
the CELIV method has been applied to the study of non-equilibrium charge
carriers photogenerated by the absorption of short laser
pulses.\cite{Juska2003,Genevicius2003} Various publications used this
photo-CELIV technique to study both mobility and density relaxation of
photogenerated
carriers.\cite{Juska2004,Osterbacka2004,Mozer2005,Mozer2005b,Mozer2005c} We
feel that it is important to point out that the analytic equations used to
evaluate CELIV experiments have been derived by assuming the presence of
\emph{equilibrium} carriers. Thus, bimolecular charge recombination has not
been taken into account for the calculation of the time-dependent charge
density $\rho(z,t)$. However, assuming charge recombination according to the
Langevin mechanism\cite{Langevin1903b}, the charge density at $z>l(t)$ decays as
$n(t)=n(0)(1+t/\tau_\sigma)^{-1}$, where $\tau_\sigma=\epsilon/en(0)\mu$ is the
dielectric relaxation time. This renders the evaluation of CELIV experiments
in terms of equation~(\ref{eq:3}) inaccurate and motivates to study the effects of charge
recombination by numerical
simulations. Figure~\ref{fig:RecombinationSimulation} shows the results of
numerically simulated current transients assuming
$\mu=2\times10^6$~V/m, $d=100$~nm, $\epsilon_r=3$ and $U'=2\times10^4$~V/s
while varying the initial charge density $n(0)$ between $10^{22}$~m$^{-3}$ and $10^{23}$~m$^{-3}$.
When bimolecular charge recombination is taken into account in the
simulations, a significantly reduced amount of charges is extracted, reducing
$\Delta j/j(0)$ and strongly shifting $t_\text{max}$ towards shorter
times. Thus, the apparent mobilities calculated from transients affected by
recombination using equation~(\ref{eq:3}) are expected to be higher compared
to those calculated from recombination-free transients. We analyzed this in
more detail for polymer blends of
poly[2,5-dimethoxy-1,4-phenylenevinylene-2-methoxy-5-(2-ethylhexyloxy)-1,4-phenylenevinylene]
(M3EH-PPV) with
poly[oxa-1,4-phenylene-1,2-(1-cyano)-ethylene-2,5-dioctyloxy-1,4-phenylene-1,2-(2-cyano)-ethylene-1,4-phenylene]
(CN-ether-PPV), for details of these materials see
Ref.~\onlinecite{Yin2008}. Samples were fabricated by spincoating a 1:1 blend
of these polymers from chlorobenzene solution onto precleaned and structured
ITO substrates covered by a layer of PEDOT:PSS (Clevios AI4083 obtained from
H.C.~Starck, Germany) and evaporating a 200~nm thick aluminum top
electrode. Devices were fabricated under protective nitrogen atmosphere and
encapsulated by a cover glass and two-component expoxy resin prior to
measurements under ambient
conditions. Figure~\ref{fig:CELIVTransientsExperiment} shows photo-CELIV
current transients obtained for a device with a 55~nm thick polymer layer and
an electrode area of $A=1$~mm$^2$ at a voltage slope of $U'=1.06$~V/$\mu$s for
various delay times $t_\text{d}$ between photogeneration using a 20~ns long
laser pulse of 355~nm wavelength and the beginning of charge extraction.
As is obvious from the current transients, the time $t_\text{max}$ of maximum
extraction current strongly shifts to smaller values for short
$t_\text{d}$. We modeled the impact of charge density on these measurement
results by numerical simulation of the experiment using the model parameters
$d=55$~nm, $U'=1.06$~V/$\mu$s, $\epsilon=3$ and assuming a field- and
time-\emph{independent} mobility $\mu=3.8\times10^{-6}$~cm$^2$/Vs. The charge
density $n(0)$ at the beginning of charge extraction was varied between
$10^{21}$~m$^{-3}$ and
$10^{25}$~m$^{-3}$. Figure~\ref{fig:RecombinationSimulationII} shows the
$\Delta j/j(0)$ and apparent mobilities calculated from the simulated CELIV
transients using equation~(\ref{eq:3}).
The apparent mobility (solid symbols) rises with charge density for
$n(0)>10^{23}$~m$^{-3}$, corresponding to current maxima (open symbols) of
$\Delta j/j(0)>0.1$. It has been suggested\cite{Genevicius2002} that CELIV transients are most
convenient to determine experimentally when $\Delta
j/j(0)\approx1$. Our simulations strongly discourage this
choice whenever nonequilibrium charge carriers are investigated, since charge
recombination strongly distorts the transients in this regime. The
simulation data can also be calculated by varying the delay time $t_\text{d}$
between photogeneration of charge carriers of density $n_\text{photo}$ and
charge extraction, whereby the charge density at the beginning of charge
extraction is $n(0)=n_\text{photo}/(1+e\mu
n_\text{photo}t/\epsilon)$. Figure~\ref{fig:CELIVExperimentEvaluation}
compares the apparent mobilities calculated from simulation data using
equation~(\ref{eq:3}) with those obtained in the same way from
figure~\ref{fig:CELIVTransientsExperiment}.
Under the assumption of $n_\text{photo}=4\times10^{23}$~m$^{-3}$, simulated
results closely follow those obtained from the measurement over the whole
range of delay times. For the shortest delay time, we estimate from the
experimental current transient that a total amount of charges corresponding to
a charge density of only $5\times10^{22}$~m$^{-3}$ could be extracted during
the CELIV experiments. For the simulated transients at a comparable delay time
of $t_\text{d}=0.38$~$\mu$s we calculated a very similar value of
$6\times10^{22}$~m$^{-3}$. Thus, more than 80\% of the initially generated
charge pairs are lost due to bimolecular recombination during the delay time
and the initial part of the CELIV transient.

\begin{figure}
  \centering
  \includegraphics{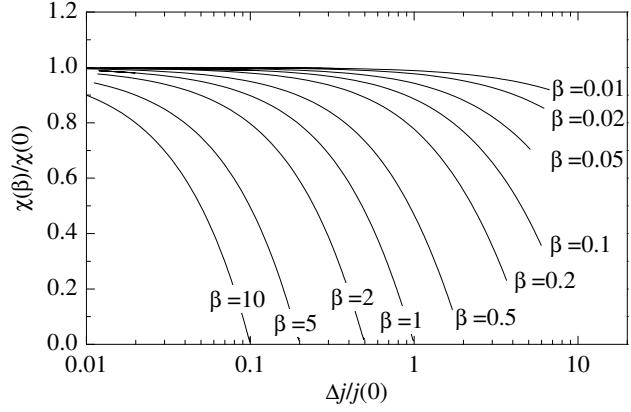}
  \caption{Results of the calculation of $\chi(\beta)/\chi(0)$ as function of
    $\Delta j/j(0)$ from the numerical solution of equations (\ref{eq:7}) and
    (\ref{eq:8}), parametric in the bimolecular recombination prefactor
    $\beta$.}
  \label{fig:RecombinationCorrectionFactor}
\end{figure}

Since the effect of charge recombination during non-equilibrium CELIV
experiments is of considerable experimental interest, we will elaborate
further on an analytic treatment of this situation. In general, the charge
density $\rho$ will follow a complicated spatial and time dependence, since 
charge recombination will take place only in the region $z>l(t)$, where both charge
types are present. To keep our analysis sufficiently general, we consider the
case of a dielectric relaxation time given by
$\tau_\sigma=\epsilon/en(0)\mu\beta$, where $\beta$ is a recombination
prefactor ($\beta=1$ corresponds to Langevin recombination). Reduced
bimolecular recombination with $\beta\ll 1$ has been shown to prevail in some
polymer/small molecule donor/acceptor blend solar cell
materials.\cite{Pivrikas2007} The spatial charge distribution for $z\leq l(t)$
then depends on $l(t)$ as
\begin{equation}
  \label{eq:9}
  \rho(z)=\frac{-en_0}{1+\tau_\sigma^{-1}l^{(-1)}(z)},
\end{equation}
where $l(l^{(-1)}(z))=z$ defines the inverse function $l^{(-1)}(z)$ of
$l(t)$. Despite this implicit expression for the charge density, analysis
shows that the current transient can be obtained by the surprisingly simple
expression
\begin{equation}
  \label{eq:7}
  j=\frac{\epsilon}{d}U'+en\frac{1-l/d}{1+t/\tau_\sigma}\frac{dl(t)}{dt}.
\end{equation}
The time dependence of $l(t)$ is calculated from
\begin{equation}
  \label{eq:8}
  \frac{d^2l(t)}{dt^2}=\frac{\mu U'}{d}-\frac{1}{d\beta\tau_\sigma}\frac{l(t)}{1+t/\tau_\sigma}\frac{dl(t)}{dt},
\end{equation}
the solution of which converges to that of equation~(\ref{eq:2}) for
$\beta\rightarrow0$. We compared this solution to the results of numerical
simulations and found good agreement when charge diffusion was neglected in
the simulations. Using the same techniques as for figure~\ref{fig:Uptmaxfit},
we calculated $\chi$ as a function of $\Delta j/j(0)$ parametric in the
prefactor $\beta$. Figure~\ref{fig:RecombinationCorrectionFactor} plots the
results relative to $\chi$ as determined for $\beta=0$.
Given a specific recombination
prefactor $\beta$, these curves can be used to directly extract the
recombination-corrected charge mobility from the measurement or to estimate
the impact of recombination on CELIV results obtained by the more simple
estimates provided by equations~(\ref{eq:6}) and~(\ref{eq:3}). We additionally
fitted $\chi$ as a function of $\Delta j/j(0)$ for the case $\beta=1$ by the double exponential
expression used above. This directly results in the expression
\begin{equation}
  \label{eq:10}
  \mu=\frac{2d^2}{U't_\text{max}^2}\left(0.860e^{-0.486\frac{\Delta j}{j(0)}}-0.525e^{0.0077\frac{\Delta j}{j(0)}}\right)
\end{equation}
which is valid at $\Delta j/j(0)<0.95$ with a relative error of less than 3.5\%
and can be used to determine true charge mobilities from CELIV transients even
under conditions of high charge densities, assuming that Langevin
recombination prevails.

\section{Field-dependent charge mobilities}
\begin{figure}
  \centering
  \includegraphics{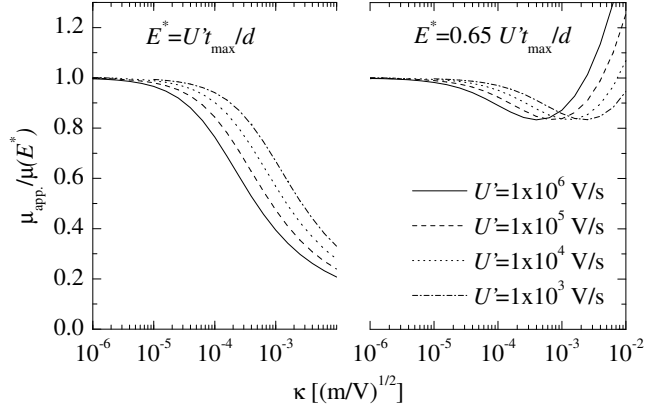}
  \caption{Ratio of the apparent mobility $\mu_\text{app.}$ and the actual
    mobility at the electric field of $E^*=U't_\text{max}/d$ (left) and
    $E^*=0.65\cdot U't_\text{max}/d$ (right) as function of the Poole-Frenkel
    parameter $\kappa$. The apparent mobility was calculated from the current
    transients given by equation~(\ref{eq:4}) using $\mu_0=10^{-6}$~cm$^2$/Vs,
    $d=100$~nm and $\epsilon_\text{r}=3$ by applying equation~(\ref{eq:6}) and is shown
    parametric in the voltage slope $U'$.}
  \label{fig:MuBetaPlot}
\end{figure}

\begin{figure}
  \centering
  \includegraphics{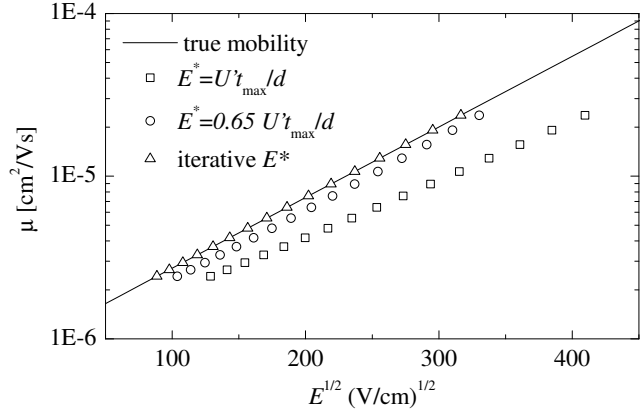}
  \caption{Apparent charge mobilities calculated from simulated CELIV
    transients using equation~(\ref{eq:6}) for field dependent charge
    mobilities as indicated by the solid line. The extraction field associated
    to determined mobility values was calculated as $E^*=U't_\text{max}/d$
    (squares), $E^*=0.65\cdot U't_\text{max}/d$ (circles) or using the iterative
    procedure as described in the text (triangles).}
  \label{fig:OptimizedTime}
\end{figure}

The charge mobility in organic semiconductors is usually considered to be both
a field and density dependent quantity, where the electric field dependence
has been experimentally found to mostly follow a Poole-Frenkel type law
$\mu(E)=\mu_0\exp(\kappa\sqrt{E})$. The work of B\"{a}ssler et al. has
shown that this can be understood in terms of transport sites having random
energies according to a gaussian distribution, rendering $\mu_0$ and $\kappa$
temperature dependent.\cite{Bassler1993} CELIV experiments provide a unique
opportunity to determine charge mobilities in undoped organic semiconductor
films of well below 100~nm thickness, but has the disadvantage of working
under conditions of a non-constant electric field. Since the mobility is
calculated from the maximum extraction current point, the determined values
have usually been associated with the electric field $E^*=U't_\text{max}/d$
(\emph{extraction field}) present at the time of maximum extraction
current,\cite{Juska2002,Mozer2005,Mozer2005c} although the validity of this
approach has never been tested rigorously. As we have shown above, mobilities
determined for non-equilibrium charge carriers using photo-CELIV are more
reliable when $\Delta j/j(0)\ll 1$, i.e. when $\tau_\sigma\gg t_\text{tr}$
where $t_\text{tr}=d\sqrt{2/\mu U'}$ is the charge transit time through the
layer. Under this approximation, the current density~(\ref{eq:1}) becomes
\begin{equation}
  \label{eq:4}
  j=\frac{\epsilon}{d}U'+
  \frac{en}{d}\left(1-\frac{l}{d}\right)U't\mu_0e^{\kappa^*}
\end{equation}
for $l(t)\leq d$, where
\begin{equation}
  \label{eq:5}
  l(t)=\frac{2\mu_0}{U'\kappa^4}\left[6d+6de^{\kappa^*}(\kappa^*-1)+U't\kappa^2e^{\kappa^*}\left(\kappa^*-3\right)\right],
\end{equation}
and $\kappa^*(t)=\kappa\sqrt{U't/d}$. Unfortunately, equation~(\ref{eq:4}) does not
provide any closed analytic expression for $t_\text{max}$ and $\Delta j/j(0)$,
but can be evaluated numerically. Figure~\ref{fig:MuBetaPlot} compares the
apparent charge mobility calculated from such data using equation~(\ref{eq:6}) to the
actual mobility at $E=E^*=U't_\text{max}/d$. 
These results were calculated for $\mu_0=10^{-6}$~cm$^2$/Vs, $d=100$~nm and $\epsilon_\text{r}=3$, but are
considered to be fairly general since they are independent of $n$ at
sufficiently low densities and invariant under the mutual transformation
$\mu_0\rightarrow\alpha\mu_0$, $U'\rightarrow\alpha U'$ for arbitrary
$\alpha$. It is obvious that significant errors in the apparent mobility occur
at large $\kappa$ and large $U'$. We propose a simple improvement of the CELIV
analysis by attributing the apparent mobility values to the extraction field
redefined as $E^*=0.65\cdot U't_\text{max}/d$. The corresponding relative
error of the apparent mobility is also shown in figure~\ref{fig:MuBetaPlot}
and stays within 20\% in the relevant parameter regime.  In order to test this
approach, we numerically simulated CELIV transients in the $\Delta j/j(0)\ll
1$ regime for various values of $U'$, assuming $d=100$~nm,
$\epsilon_\text{r}=3$, $\mu_0=10^{-6}$~cm$^2$/Vs and
$\kappa=10^{-3}$~(m/V)$^{1/2}$. Figure~\ref{fig:OptimizedTime} compares the
apparent mobility values determined according to equation~(\ref{eq:6}) and
associated with either choice of $E^*$.
We found that using $E^*=0.65\cdot U't_\text{max}/d$ generally gives better results
for this type of field dependence. The experimental error can be further
minimized by using an iterative procedure if the field-dependence of the apparent
mobility indeed follows the Poole-Frenkel behaviour: (1.) measure CELIV
transients at different $U'$ to obtain a range of $\mu$ and
$t_\text{max}$ values via equation~(\ref{eq:6}), (2.) determine preliminary parameters $\mu_0^{(0)}$ and
$\kappa^{(0)}$ from the measurement using $E^*=U't_\text{max}/d$, (3.) for each
U', calculate the theoretical CELIV transient from equation~(\ref{eq:4}),
numerically evaluate $(\Delta j/j(0))^\text{th}$, $t_\text{max}^\text{th}$ and $\mu^\text{th}$
using equation~(\ref{eq:6}) and calculate $\delta=d\ln(\mu^\text{th}/\mu_0^{(0)})/U't_\text{max}^\text{th}\kappa^{(0)}$,
(4.) associate each \emph{measured} mobility value to the extraction field
$E^*=\delta U't_\text{max}/d$ and
determine optimized $\mu_0^{(1)}$ and $\kappa^{(1)}$ from this data, (5.) iterate the
procedure by repeating steps 3 and 4 until the determined $\mu_0^{(n)}$ and $\kappa^{(n)}$
stabilize. This procedure is only moderately complex but significantly
enhances the accuracy of charge mobility determination, at least when
Poole-Frenkel field dependence prevails. Figure~\ref{fig:OptimizedTime}
shows that the results of this iteration procedure accurately track the true
field-dependent mobility used for the simulated CELIV experiments.

\section{Conclusion}
\label{sec:conclusion}

In this publication, we pointed out several difficulties that arise when
applying the CELIV technique under the non-idealized conditions usually
encountered in experiments. We based our analysis upon a rederivation of the
original CELIV analysis, correcting for inaccuracies in the original
publications that result in erroneous charge mobilities under conditions of
high charge density. In the case of photogenerated charge carriers, we found a
significant departure from the equilibrium assumption of the original
derivations.  The impact of bimolecular charge recombination on CELIV
transients and their analysis in this situation has not been considered up to now. We showed
that high charge densities as typically used in the experiments leads to an
artifical time dependence of the determined mobility values. We were able to
relate the experimental results for a M3EH-PPV/CN-ether-PPV blend solar cell
solely to this effect, showing that the charge mobility is actually constant
shortly after photoexcitation. Sufficient information was provided to
facilitate an interpretation of experiments under conditions of
non-equilibrium charge carrier extraction. As another strong deviation from
idealized conditions we investigated the effect of field-dependent charge carrier
mobilities. We showed that association of the CELIV mobilities with the
electric field present at the time of extraction current maximum leads to significant
errors in the determined field dependence. An optimized choice of the
correlated extraction field was introduced, which yields much lower errors
compared to the standard approach. Additionally, we showed that under the
presumption of a Poole-Frenkel type field dependence, an iterative procedure
can be applied to determine the true mobility-field dependence in a precise
way.

\begin{acknowledgments}
S.\ B.\ acknowledges financial support by the 
German Federal Ministry of Science and Education (BMBF
FKZ 13N8953 and 03X3525D). M.\ Sch.\ acknowledges financial support by
the German Research Foundation (DFG SPP 1355). 
\end{acknowledgments}

\end{document}